\documentclass[sigconf]{acmart}

\usepackage{graphicx}
\usepackage{array}
\usepackage{booktabs}

\newcommand{\colimgwidth}{\columnwidth}
\newcommand{\fullimgwidth}{\textwidth}

\AtBeginDocument{%
  }

\copyrightyear{2025} 
\acmYear{2025} 
\setcopyright{cc}
\setcctype{by}
\acmConference[ISCA '25]{Proceedings of the 52nd Annual International Symposium on Computer Architecture}{June 21--25, 2025}{Tokyo, Japan}
\acmBooktitle{Proceedings of the 52nd Annual International Symposium on
Computer Architecture (ISCA '25), June 21--25, 2025, Tokyo, Japan}
\acmDOI{10.1145/3695053.3731006}
\acmISBN{979-8-4007-1261-6/2025/06}

\begin{document}

\title{LightNobel: Improving Sequence Length Limitation in Protein Structure Prediction Model via Adaptive Activation Quantization}


\author{Seunghee Han}
\email{shhan1755@kaist.ac.kr}
\authornote{Both authors contributed equally to this research.}
\orcid{0009-0009-0953-028X}
\affiliation{%
  \institution{KAIST}
  \city{Daejeon}
  \country{South Korea}
}

\author{Soongyu Choi}
\authornotemark[1]
\email{soongyu1291@kaist.ac.kr}
\orcid{0009-0005-6453-9755}
\affiliation{%
  \institution{KAIST}
  \city{Daejeon}
  \country{South Korea}
}

\author{Joo-Young Kim}
\email{jooyoung1203@kaist.ac.kr}
\orcid{0000-0003-1099-1496}
\affiliation{%
  \institution{KAIST}
  \city{Daejeon}
  \country{South Korea}
}


\begin{abstract}

Recent advances in Protein Structure Prediction Models (PPMs), such as AlphaFold2 and ESMFold, have revolutionized computational biology by achieving unprecedented accuracy in predicting three-dimensional protein folding structures. However, these models face significant scalability challenges, particularly when processing proteins with long amino acid sequences (e.g., sequence length > 1,000). The primary bottleneck that arises from the exponential growth in activation sizes is driven by the unique data structure in PPM, which introduces an additional dimension that leads to substantial memory and computational demands. These limitations have hindered the effective scaling of PPM for real-world applications, such as analyzing large proteins or complex multimers with critical biological and pharmaceutical relevance.

In this paper, we present LightNobel, the first hardware-software co-designed accelerator developed to overcome scalability limitations on the sequence length in PPM. At the software level, we propose Token-wise Adaptive Activation Quantization (AAQ), which leverages unique token-wise characteristics, such as distogram patterns in PPM activations, to enable fine-grained quantization techniques without compromising accuracy. At the hardware level, LightNobel integrates the multi-precision reconfigurable matrix processing unit (RMPU) and versatile vector processing unit (VVPU) to enable the efficient execution of AAQ. Through these innovations, LightNobel achieves up to 8.44$\times$, 8.41$\times$ speedup and 37.29$\times$, 43.35$\times$ higher power efficiency over the latest NVIDIA A100 and H100 GPUs, respectively, while maintaining negligible accuracy loss. It also reduces the peak memory requirement up to 120.05$\times$ in PPM, enabling scalable processing for proteins with long sequences.

\end{abstract}

\ccsdesc[500]{Hardware~Hardware accelerators}
\ccsdesc[500]{Hardware~Application specific integrated circuits}
\ccsdesc[300]{Computer systems organization~Neural networks}
\ccsdesc[100]{Applied computing~Computational biology}

\keywords{Protein Structure Prediction Model (PPM), AlphaFold2, ESMFold, Attention-based Model, Quantization, Hardware Accelerator, \newline Hardware-Software Co-design}




\maketitle

\section{Introduction}
\label{section_1}

\begin{figure}[t!]
\centering
\includegraphics[width=\colimgwidth]{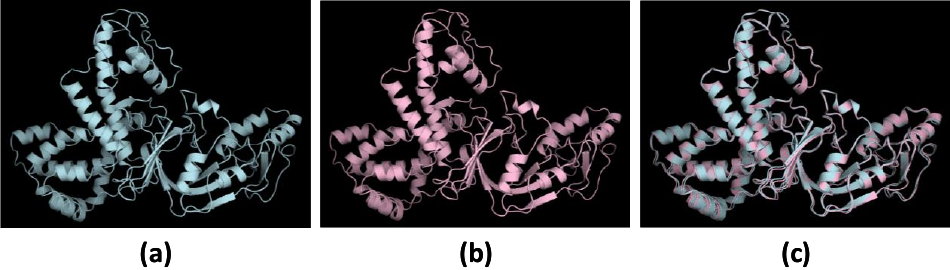}
\caption{Visualization of Protein Structure Prediction Model (PPM) results. (a) Result of the conventional PPM. (b) Result of LightNobel. (c) Comparison between the two results.}
\label{fig_1_1}
\end{figure}

Proteins, the fundamental building blocks of life, are at the center of nearly all biological processes. These essential biomolecules, composed of amino acid sequences, fold into intricate three-dimensional structures that dictate their functions~\cite{protein_3d_struct}. An accurate understanding of protein folding structures is critical for deciphering their roles in protein binding, interactions, and other functions essential to biotechnological and pharmaceutical applications.

However, accurately determining protein folding structures remains a formidable challenge.
Experimental methods~\cite{X_ray_cry, cryo-electron} have historically been used to explore these structures.
While effective, these approaches are costly and time-consuming. Over the past 60 years, only 170,000 protein structures have been identified, but the total number of proteins across all known living organisms is estimated to exceed 200 million~\cite{protein_num}. This disparity underscores the urgent need for more efficient and scalable approaches.

The advent of computational models marked a turning point in the Protein Structure Prediction Model (PPM). However, they struggled to achieve the accuracy required to replace experimental methods at the beginning. At this juncture, the application of deep learning to protein folding prediction has dramatically transformed this field.
Early models~\cite{Alphafold1, RaptorX-Contact} leveraged Convolutional Neural Networks (CNNs)~\cite{CNN}, delivering substantial improvements over conventional approaches.
The introduction of attention-based models~\cite{Attention} further advanced the field by considering interactions across all positions in the input sequence. AlphaFold2~\cite{AlphaFold2} integrated a large-scale database with an attention-based model, achieving unprecedented accuracy in CASP14~\cite{CASP14}, the premier competition in protein structure prediction. ESMFold~\cite{ESMFold} enhanced prediction speed by leveraging ESM-2, a type of protein language model, instead of relying on extensive databases. These advances culminated in the awarding of the 2024 Nobel Prize in Chemistry~\cite{Nobel_prize}.

Despite their impressive performance, PPM faces significant scalability challenges. 
The Pair Representation in the attention-based PPM introduces an additional dimension, which significantly inflates memory requirements as sequence length increases. This special data structure not only limits scalability to long protein sequences but also results in high latency.
However, the need to analyze long-sequence proteins is becoming increasingly urgent.
For example, long proteins such as titin~\cite{tinin}, which play diverse and vital roles, consist of tens of thousands of amino acids.
Additionally, proteins frequently form complexes, or multimers, inherently increasing sequence length to perform biological functions. This demand is evident in CASP competition~\cite{CASP}, where targeting sequence lengths have grown from 770 in CASP10 to 6,879 in CASP16, emphasizing the need for scalable solutions. Some systems, such as OpenFold~\cite{OpenFold}, address this demand through techniques such as chunking and Low-Memory Attention (LMA)~\cite{LMA}, extending support to sequences of up to 4,600 amino acids. Nonetheless, these methods remain insufficient for handling the rapidly increasing sequence lengths of protein structure demands.

To address these challenges, we focus on quantization.
While quantization is widely applied in attention-based models such as LLMs~\cite{LLM} or ViTs~\cite{ViT}, they mainly focus on weights as the size of weights represents a major bottleneck in such models~\cite{AWQ, BiLLM, GOBO, MiniViT}.
Some studies deal with simultaneous quantization of both weights and activations, but they are applied in a limited manner and show accuracy loss due to the limitations of quantization~\cite{SmoothQuant, QServe, AffineQuant, Mokey, LLM_INT8}. However, in PPM, maintaining prediction accuracy is critical, as precise folding results are essential for practical application.

Through detailed analysis, we discover that each activation in PPM exhibits unique characteristics.
This insight enables us to design a new approach that dynamically adjusts precision and outlier handling based on the characteristics of different activations. 
Additionally, the hardware-unfriendly multi-precision and dynamic dataflow is solved through our special hardware design.
Figure~\ref{fig_1_1} demonstrates that our system achieves nearly identical results to the conventional system in protein folding prediction.

In this paper, we propose LightNobel, a hardware-software co-designed solution that overcomes the scalability limitations on the sequence length due to the activation size through a novel quantization approach combined with meticulously designed dedicated hardware.
The main contributions of our work are as follows.
\begin{itemize}
\item We identify the severe limitations of the Protein Structure Prediction Model (PPM) in handling long sequences, primarily due to rapidly increasing activation size, which leads to high peak memory requirements and latency.
\item We develop a Token-wise Adaptive Activation Quantization (AAQ) scheme that analyzes the token-wise characteristics of activations in the PPM and applies precision and outlier handling differently, suggesting a new approach to solve the activation size problem without accuracy degradation.
\item Based on AAQ, we propose LightNobel, a hardware accelerator that flexibly supports multi-precision and dynamic outlier configuration among quantized activations, and configures a dataflow that can maximize hardware utilization for different types of activations.
\item Our experiments show LightNobel achieves up to 8.44$\times$, 8.41$\times$ speedup and 37.29$\times$, 43.35$\times$ higher power efficiency over the latest NVIDIA A100 and H100 GPUs while maintaining negligible accuracy loss and reducing peak memory requirements by 120.05$\times$ for proteins with long sequences.
\end{itemize}

\section{Background}
\label{section_2}

\subsection{Attention-based Model}
\label{section_2_1}

Attention-based models are models with attention layers, widely used for various applications~\cite{LLM, ViT}.
The attention layer operates in the following steps. First, Query (Q), Key (K), and Value (V) are computed from the input. Next, attention scores are calculated through dot product operations between Q and K, followed by scaling.
After applying a softmax function, the resulting attention weights reflect the importance of each element. This step captures the underlying relationships in the data.
Finally, the output is obtained by multiplying these weights with V, typically followed by linear layers.
In the case of Multi-head Attention (MHA), QKV generation and attention score calculation can be divided into multiple groups called heads.
Attention-based models have achieved remarkable success across various fields, owing to their applicability and scalability to numerous tasks. Their ability to consider interactions across all positions in a sequence has recently extended their application to the Protein Structure Prediction Model (PPM).

\subsection{Model Quantization}
\label{section_2_2}

Model Quantization is a technique that represents high-precision values (e.g., FP32) as discrete lower-precision values (e.g., INT4, INT8). With the increasing scale and complexity of modern deep learning models, especially attention-based models, quantization has gained significant attention for its ability to reduce computational costs and memory requirements~\cite{quant_nn}.

\textbf{Scaling Factor.} Since the representation range of lower precision values is narrower compared to higher precision, scaling is required to represent higher precision values.
The multiplier used in this scaling is called the scaling factor.
Scaling factors are predetermined using calibration or are dynamically adjusted at runtime for quantization and further utilized for dequantization.

\textbf{Quantitation Granularity.} If every value shares the same scaling factor, quantization accuracy significantly decreases due to poor representation of values.
To address this, values can be divided into groups with small variances and share the same scaling factor, called quantization granularity, which defines the level of grouping.
Quantization applied to the tensor is called tensor-wise quantization, the channel-level group is called channel-wise quantization, and the token-level group is called token-wise quantization.

\textbf{Outlier Handling.} One critical challenge in the quantization process is outlier handling. Outliers are values significantly larger or smaller than the mean within a quantization granularity group, disrupt the small variance among values, and affect quantization accuracy. To address this, distinguishing outliers and handling them separately is necessary.

\begin{figure*}[t]
\centering
\includegraphics[width=\fullimgwidth]{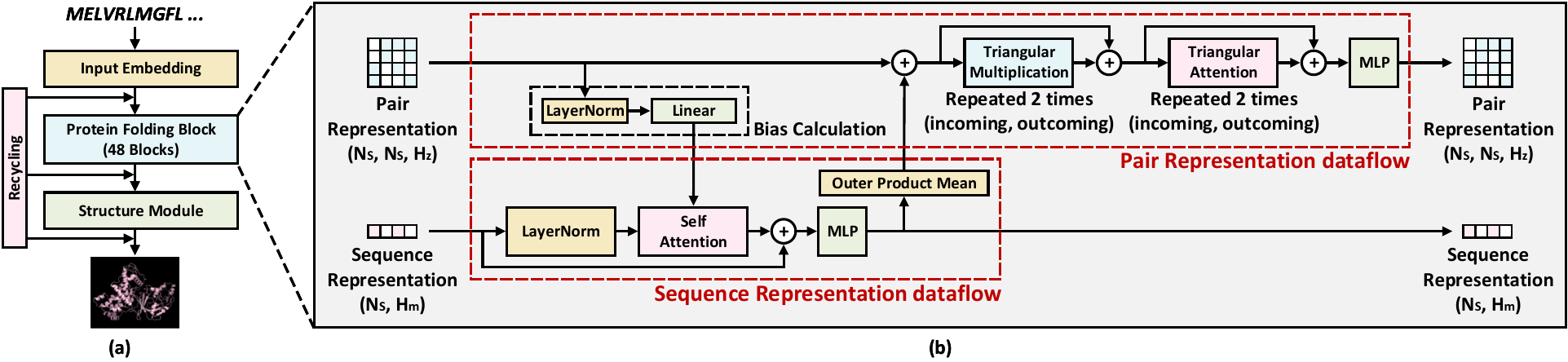}
\caption{Overview of PPM. (a) Block diagram of the PPM. (b) Dataflow of Protein Folding Block (1 Block).}
\label{fig_2_2}
\vspace{0.10in}
\end{figure*}

\subsection{Protein Structure Prediction Model (PPM)}
\label{section_2_3}

Protein Structure Prediction Model (PPM) aims to predict the three-dimensional folding structure of a protein from its amino acid sequence.
Recent state-of-the-art PPMs~\cite{AlphaFold2, ESMFold, OpenFold, RoseTTAFold} demonstrate exceptional performance through the use of attention mechanisms.

Figure~\ref{fig_2_2}(a) shows the overall process of PPM, including Input Embedding, Protein Folding Block, and the Structure Module. In the Input Embedding stage, the amino acid sequence of a protein is taken as input and converted into two types of biological information.
Pair Representation contains information about interactions between pairs of amino acids in the protein sequence.
It undergoes iterative updates to capture the protein's distogram patterns, reflecting positional relationships. Pair Representation has a dimension of $(\mathrm{Ns}, \mathrm{Ns}, \mathrm{Hz})$, where $\mathrm{Ns}$ is the length of the protein sequence and $\mathrm{Hz}$ is the hidden dimension of Pair Representation, typically set to 128, which is a significantly smaller value compared to other attention-based models~\cite{LLaMA}. 
Here, a token in PPM means a vector in the $\mathrm{Hz}$ direction with a dimension of $(\mathrm{1}, \mathrm{1}, \mathrm{Hz})$, similar to attention-based models~\cite{SmoothQuant}.
Sequence Representation contains information derived from other organisms similar to the input protein. It has a dimension of $(\mathrm{Ns}, \mathrm{Hm})$, where $\mathrm{Hm}$ is the hidden dimension of Sequence Representation, typically set to 1024. Some models, such as AlphaFold2~\cite{AlphaFold2}, use Multiple Sequence Alignment (MSA) as the Sequence Representation, combining information from multiple organisms.

In the Protein Folding Block stage, the core attention mechanism is applied. At this stage, both Pair and Sequence Representations are used together to capture relationships between all sequence positions in the amino acid sequence, forming the information required for protein structure prediction.
The Evoformer in AlphaFold2~\cite{AlphaFold2} and the Folding trunk in ESMFold~\cite{ESMFold} are examples of this block. Figure~\ref{fig_2_2}(b) shows the dataflow of a Protein Folding Block, especially a Folding trunk in ESMFold.
A key part of this stage is the Pair Representation dataflow, which effectively captures the interactions between amino acid positions.
In the Triangular Multiplication block, interaction patterns between amino acids are refined, enabling more precise learning of protein structure prediction.
In the Triangular Attention block, attention mechanisms are applied to fine-tune the relationships between amino acid positions while updating the Pair Representation.
Details of these dataflows are provided in Figure~\ref{fig_4_2}(a) and (b).
In the Structure Module stage, the Pair Representation generated in the previous stage is used to predict the actual three-dimensional structure of the protein.
To improve prediction accuracy, a recycling process is employed to iteratively refine the predicted results.

\subsection{TM-Score}

TM-Score is a metric widely used in structural biology and computational biology to evaluate the similarity between predicted and actual three-dimensional protein structures. TM-Score measures the global similarity between two structures, providing a stable evaluation criterion even when comparing proteins of different sizes.
TM-Score plays a critical role in assessing the quality of predicted protein models and comparing the performance of structure prediction algorithms. The TM-Score ranges from 0 to 1, with values closer to 1 indicating higher similarity between the two structures. Generally, a TM-Score of 0.5 or higher is considered indicative of strong structural similarity~\cite{TM-align}. TM-Score has become a standard metric in the field of protein prediction and is widely used as a reliable indicator for evaluating the accuracy of PPM.

\section{Motivation}
\label{section_3}

\subsection{PPM Latency Analysis}
\label{section_3_1}

\begin{figure}[t]
\centering
\includegraphics[width=\colimgwidth]{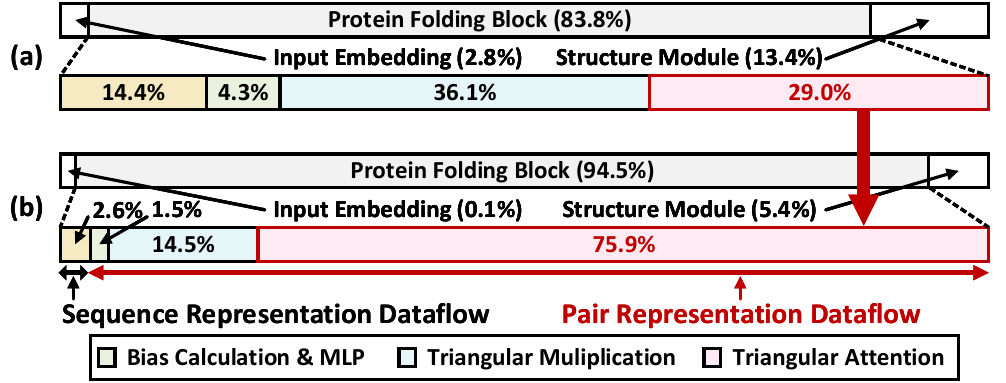}
\caption{Latency Breakdown of PPM with (a) protein R0271 (77 amino acids) and (b) protein T1269 (1,410 amino acids).}
\label{fig_3_1}
\end{figure}

To identify the bottleneck impacting the performance of PPM, we conduct an end-to-end latency breakdown of the entire execution time.
Figure~\ref{fig_3_1} shows the latency breakdown when running the PPM.
In this experiment, we use ESMFold~\cite{ESMFold}, which significantly improves execution speed compared to AlphaFold2~\cite{AlphaFold2} by leveraging an additional language model for Input Embedding.
All experiments are conducted on an NVIDIA H100 GPU~\cite{H100} using the vanilla model without the chunk option.
For the dataset, we select proteins from the latest CASP16~\cite{CASP16} dataset. Specifically, we select R0271 (77 amino acids) as the shortest protein and T1269 (1,410 amino acids) as the longest protein that can be processed within a single GPU.

As shown in Figure~\ref{fig_3_1}, the Protein Folding Block in the PPM constitutes a significant portion of the total execution time, accounting for 83.8\% and 94.5\%, respectively.
The Pair Representation dataflow accounts for 69.4\% of the total execution time for R0271 as shown in Figure~\ref{fig_3_1} (a), but the proportion increases dramatically to 91.9\% for T1269 as shown in Figure~\ref{fig_3_1} (b), indicating that it becomes a major bottleneck. 
This sharp increase is primarily due to the execution time spent on the Triangular Attention operation, which surges from 29.0\% to 75.9\%.
As mentioned in Section~\ref{section_2_3}, the size of the Pair Representation grows quadratically with the sequence length, unlike other data structures.
Consequently, as the sequence length increases, the proportion of Pair Representation dataflow within the overall PPM execution grows significantly. For extremely large proteins such as PKZILLA-1 (45,212 amino acids)~\cite{PKZILLA}, Pair Representation dataflow is expected to account for more than 99\% of the total PPM runtime.

\begin{figure}[t]
\centering
\includegraphics[width=\colimgwidth]{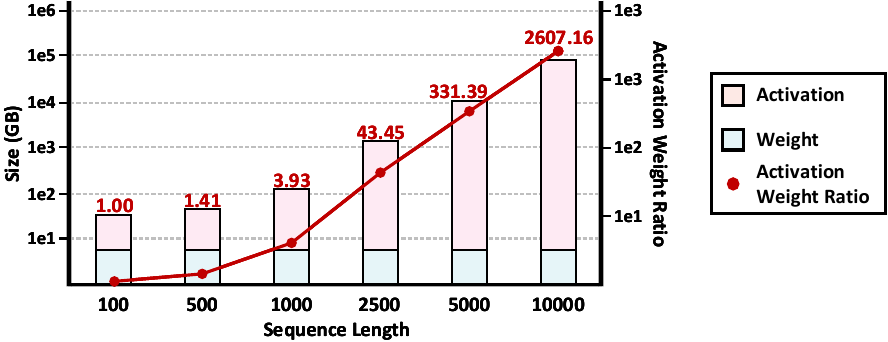}
\caption{Analysis of total weight size and peak activation size in PPM across various sequence lengths.}
\label{fig_3_2}
\end{figure}

\subsection{Activation Size Explosion}
\label{section_3_2}

Typical attention-based models involve major activations such as input, Q, K, and V, which have dimensions of $(\mathrm{Ns}, \mathrm{H})$, while the attention score matrix for each head has dimensions of  $(\mathrm{Ns}, \mathrm{Ns})$, where $\mathrm{Ns}$ is sequence length and $\mathrm{H}$ is the hidden dimension.
In contrast, as mentioned in Section~\ref{section_2_2}, the main bottleneck of the PPM, Pair Representation dataflow, involves major activations with dimensions of $(\mathrm{Ns}, \mathrm{Ns}, \mathrm{Hz})$, and its attention score matrix for each head has dimensions of $(\mathrm{Ns}, \mathrm{Ns}, \mathrm{Ns})$.
As a result, the memory footprint and computation cost scale cubically with sequence length in score matrix operations and quadratically for others. 
Although the hidden dimension is relatively small compared to typical models, this difference does not alleviate the bottleneck.

Figure~\ref{fig_3_2} shows the analysis of weight and peak activation size in the PPM across varying sequence lengths. As shown in the figure, the activation size increases significantly with sequence length and is very large compared to the weight size. 
At a sequence length of 2034, the activation size is already 24.15$\times$ larger than the weight size, requiring 144 GB of memory, which exceeds the capacity of a single state-of-the-art GPU~\cite{H200}. 
This exponential growth in activation size not only increases memory footprint and computation demands but also decelerates model inference and significantly raises peak memory requirements. 
While data chunking can reduce peak memory requirements by splitting computations, it is not a fundamental solution since it substantially increases the memory footprint, causing significant performance degradation.

We propose applying quantization to address the above challenge. Quantization can serve as a fundamental solution, as it reduces the original data size, thereby reducing both the memory footprint and compute workload. We further propose leaving the weights unquantized to preserve their information density, focusing instead on activation quantization, since the size of the weights is significantly smaller than that of the activations. This approach aims to maximize the quantization efficiency while minimizing accuracy degradation. Thus, we introduce an activation-only quantization scheme tailored for PPM.

\begin{figure}[t]
\centering
\includegraphics[width=\colimgwidth]{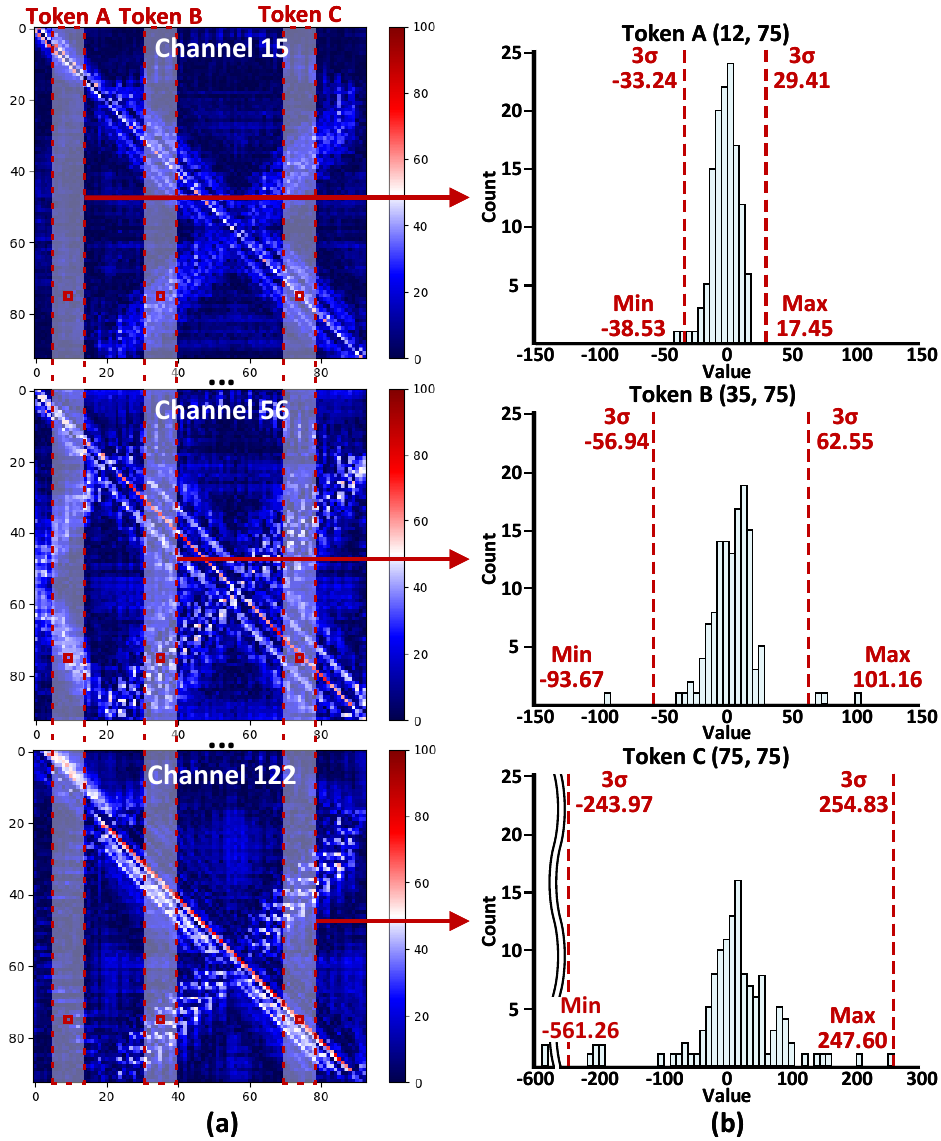}
\caption{Analysis of activation value distribution in PPM. Visualization of (a) three representative channels and (b) three representative tokens.}
\label{fig_3_3}
\vspace{-0.05in}
\end{figure}

\subsection{Token-Wise Distogram Pattern in Activation}
\label{section_3_3}

One of the most important considerations in quantization is maintaining model accuracy. 
Designing a suitable quantization scheme tailored to the characteristics of the data significantly impacts the model's accuracy.
Therefore, we analyze the distribution of activation values to identify the best quantization scheme for PPM.

It is widely recognized that general attention-based models exhibit a large variance between channels, which leads to channel-wise quantization~\cite{SmoothQuant}.
However, activations in PPM show a rel- atively small variance between channels and a large variance between tokens.
Figure~\ref{fig_3_3}(a) shows the distribution of absolute values for three random channels, and Figure~\ref{fig_3_3}(b) shows the value distribution of the random tokens from the same PPM activation. 
As shown in Figure~\ref{fig_3_3}(a), the variance between channels is small, as it shows a similar pattern.
However, as shown in Figure~\ref{fig_3_3}(b), the variance between tokens is large, as they show significantly different value ranges and distributions depending on their position.
All channels have similar minimum and maximum values at the same position, and outliers identified by the $3\sigma$-rule~\cite{3SIGMA, Olive} are concentrated only at tokens in certain positions.
This behavior arises because the Pair Representation in PPM shares distogram patterns specific to the input protein, which is a pairwise distance representation that captures spatial relationships specific to protein structures~\cite{distogram}.
Therefore, we chose a token-wise quantization scheme that aligns with the characteristics of tokens to minimize accuracy degradation.

\subsection{Characteristic Difference Across Activations}
\label{section_3_4}

To efficiently perform token-wise quantization in PPM, we analyze the characteristics of every activation value across all iterations.
Precise understanding of these activation patterns is essential for designing effective quantization schemes, particularly in setting appropriate scaling factors and handling outliers.
If the scaling factor is too large, quantization errors increase as the gap between original and quantized values widens. Conversely, if the scaling factor is too small, large values may get clipped or distorted due to the limited range of representation. While handling outliers can mitigate these issues, it increases data size and complicates dataflow, reducing overall efficiency. 
Therefore, a balanced design between scaling and outlier handling is important.

Prior quantizations for attention-based models have predominantly employed conservative activation quantization strategies to preserve model accuracy~\cite{SmoothQuant, Olive}.
For instance, quantization for activations such as pre-LayerNorm or Score Matrix is rarely explored.
However, to minimize activation overhead, we aim to quantize most of the activations in Pair Representation dataflow, including these underexplored regions. 
At the same time, it is essential to maintain accuracy comparable to the baseline in the PPM.
Our analysis reveals that activations within a single PPM layer exhibit diverse characteristics depending on their position, posing challenges to optimizing all activations using a single quantization scheme. Consequently, we propose an adaptive quantization approach that applies different quantization schemes based on the characteristics of each activation, thereby maximizing model performance.

\section{Token-wise Adaptive Activation Quantization}
\label{section_4}

As mentioned in Section~\ref{section_3_1}, the significantly large activation size in the Protein Structure Prediction Model (PPM) presents a major limitation. To address this, we propose Token-wise Adaptive Activation Quantization (AAQ), which optimally quantizes Pair Representation activations by fully leveraging their characteristics.
By integrating dynamically adjusted precision and adaptive outlier handling into token-wise quantization, AAQ effectively mitigates activation size issues in PPM while ensuring accurate inference.

\subsection{Baseline Token-wise Quantization}

AAQ specifically targets activations in Pair Representation dataflow, the main bottleneck of PPM as mentioned in Section~\ref{section_3_1}.
To maximize model accuracy, we use a 16-bit fixed-point format for weights without quantization. For activations, quantized inlier values are represented using INT4 or INT8 formats, and outliers are represented in an INT16 format to minimize information loss.

\textbf{Token-wise Quantization.}
In an attention-based model, including PPM, most computations, including linear layers or layer normalization, are performed token-wise.
Therefore, supporting dynamic token-level parallelism is critical for efficient processing in attention-based models~\cite{SOFA}. However, conventional attention-based models employing quantization often use channel-wise quantization for accuracy. This approach necessitates the dequantization of individual values within a token before every operation to enable token-wise parallel computations, making the process highly inefficient~\cite{SmoothQuant, Tender}.
In contrast, PPM also exhibits a pattern in which large values are concentrated in specific tokens. To exploit this property, we adopt a token-wise quantization.
By applying token-wise quantization and setting the scaling factor dynamically at runtime, where each token is adjusted with a unique scaling factor, we achieve superior quantization accuracy.

\textbf{Dynamic Outlier Handling.}
Although activations in PPM can be quantized token-wise, handling outliers remains another challenge.
In typical attention-based models, channel-wise quantization allows predetermination of quantization parameters based on dataset analysis.
However, since the number of tokens varies significantly depending on input, predefining thresholds for outlier classification is not feasible in a token-wise manner.
Additionally, in PPM, unpredictable outliers arise due to biasing and merging with Sequence Representation, depending on the input protein type, making it more difficult.
To resolve this, we propose a dynamic outlier handling.
During each quantization process, we utilize a top-k algorithm to identify the dynamic number of outliers. The number of outliers, k, can be set adaptively based on the activation characteristics. The computational complexity of the top-k selection is $O(nlogn)$, making it impractical for large-scale attention-based models. However, in PPM, the hidden dimension is just 128, which is very small compared to the general attention-based model~\cite{LLaMA}, and the cost is manageable. We further design hardware support for efficient top-k algorithms, enabling this approach.

\textbf{Uniform and Symmetric Quantization.} After outlier handling, the distribution of inliers is uniform. Hence, we employ uniform symmetric quantization for each token.
Equation~\ref{eq_4_1} presents the formulation of uniform symmetric quantization. Here, Min and Max denote the minimum and maximum values within the target quantization group, $\sigma$ represents the scaling factor, and m is the bit-width used for quantization.
 
\vspace{-0.05in}
\begin{equation}
 M = \max(|\text{Min}|, |\text{Max}|), \; \ \sigma = \frac{M}{2^{m-1} - 1}, \; \  Q(x) = \text{round} \left( \frac{x}{\sigma} \right)
\label{eq_4_1}
\end{equation}
\vspace{-0.05in}

\begin{figure*}[t]
\centering
\includegraphics[width=\fullimgwidth]{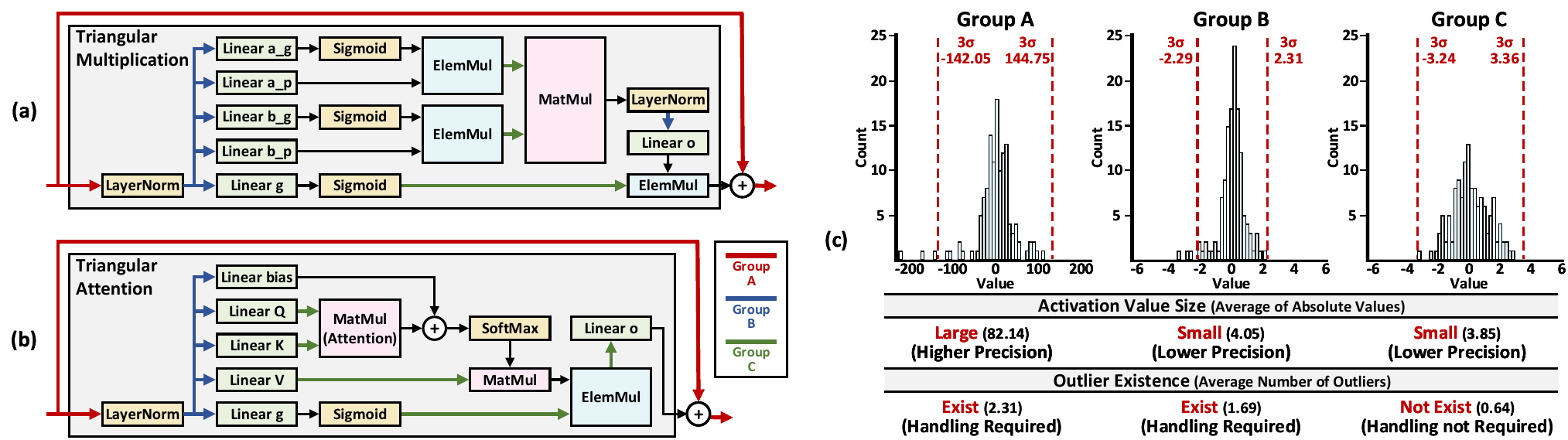}
\caption{Dataflow of (a) Triangular Multiplication Block and (b) Triangular Attention Block. Red, blue, and green lines correspond to activations belonging to each group. Black lines correspond to activations in which quantization does not occur in our dataflow. (c) Characteristic analysis of activations belonging to each group.}
\label{fig_4_2}
\end{figure*}

If the absolute difference between the scale range within a quantization group is significant, asymmetric quantization can use additional bias to focus on a narrower range. However, according to our experimental results, symmetric quantization without outlier handling leads to a 27.35\% increase in RMSE. In contrast, when outlier handling was applied, the RMSE increase is limited to only 9.76\%, corresponding to a negligible real-value difference of 0.0004. This slight difference demonstrates that symmetric quantization becomes effective when combined with outlier handling. Therefore, we adopt uniform symmetric quantization with dynamic outlier handling.

\subsection{Activation Adaptive Quantization}
\label{section_4_2}

As mentioned in Section~\ref{section_3_4}, the activations of PPM exhibit distinct characteristics, necessitating quantization schemes tailored to them.
Therefore, we adaptively optimize various quantization schemes that refine the baseline quantization scheme to suit each activation's characteristic, thereby enhancing the accuracy of quantization.

To classify activations, we focus on two key features essential for quantization: the value range and the existence of outliers.
We sample proteins from the CAMEO~\cite{CAMEO}, CASP14~\cite{CASP14}, CASP15~\cite{CASP15}, and CASP16~\cite{CASP16} datasets and analyze every token from activations. For each token, we compute the average of absolute values and the number of outliers per token.
We use the $3\sigma$-rule~\cite{3SIGMA, Olive} to identify outliers~\cite{Olive}.
Based on these features, we categorize the activations in the PPM block into three groups.
Figure~\ref{fig_4_2}(a) and (b) show which group each activation is included in the Triangular Multiplication Block dataflow and Triangular Attention Block dataflow, which are the core of the PPM block. Also, Figure~\ref{fig_4_2}(c) shows the characteristics and experiment results of activations within each group. 
First, activations in Group A are located before the LayerNorm layer and are directly connected to residual connections. 
These activations propagate data with large values and outliers through residual connections. The values vary significantly, with an average of 82.14, while having 2.31 outliers on average.
Therefore, during quantization, it is essential to allocate relatively high precision to inliers to secure a sufficient representation range and implement outlier handling. 
Activations in Group B have passed through the LayerNorm layer but have not yet undergone linear layers. Due to the normalization effect of the LayerNorm, the values are reduced compared to Group A, resulting in an average value of 4.05. However, outliers still exist in the distribution, averaging 1.69 outliers. Thus, while outlier handling remains necessary, relatively lower precision can be assigned to inliers without compromising quantization accuracy. 
Finally, Group C consists of activations that do not belong to the previous two groups. These activations undergo multiplication with relatively small weights, resulting in an average value of 3.85, comparable to Group B. Their distribution has very few outliers, having an average of 0.64 outliers, which is less than 1. Thus, satisfactory accuracy can be achieved without handling outliers during quantization.

The proposed AAQ method dynamically adjusts quantization schemes based on activation characteristics to optimize both accuracy and efficiency. 
As detailed in Section~\ref{section_7_1}, we determine the most efficient quantization schemes for each group through design space exploration. 
Based on this result, AAQ dynamically determines which values should be handled as outliers in each token at runtime, depending on the type of activation.
From the perspective of outlier handling, AAQ dynamically handles a variable number of outliers. From the perspective of inlier handling, it adopts a multi-precision approach, thereby aligning the quantization scheme with the unique characteristics of each activation. Our adaptive approach strategically aligns with activation characteristics, achieving optimal accuracy while significantly reducing memory consumption and computational overhead. From a hardware-software co-design perspective, we achieve this by implementing multi-precision dataflow for handling token-wise dynamic quantization schemes and supporting top-k functionality for dynamic outlier handling in our hardware.

\begin{figure}[t]
\centering
\includegraphics[width=\colimgwidth]{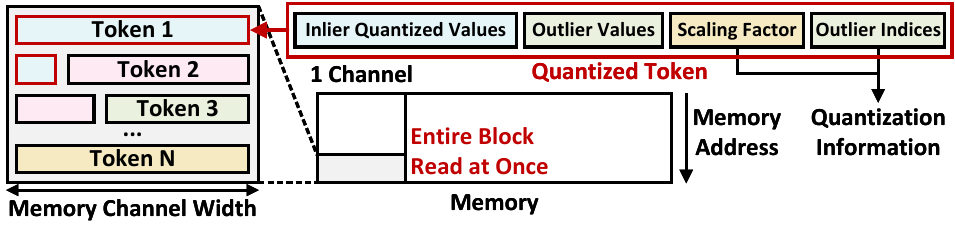}
\caption{Memory layout of the quantized tokens.}
\label{fig_4_3}
\end{figure}

\subsection{Memory Layout for Quantization}
\label{section_4_3}

To accommodate diverse quantization schemes, we implement a scalable and flexible memory mapping. Figure~\ref{fig_4_3} shows the layout of quantized tokens in the memory. Inliers are stored sequentially, followed by outliers, scaling factors, and outlier indices.
To optimize memory utilization, data from multiple tokens is grouped into blocks, with the block size determined based on the bandwidth of the memory channel.

Despite token-wise quantization sharing the scaling factor on a token level, inliers require a dequantization process in the last step of computation, while outliers do not. Therefore, inlier-outlier adjacency minimizes scaling factor computations, enhancing hardware utilization.
This out-of-order value mapping is feasible due to our hardware support for matrix multiplication without dequanti- zation during runtime, particularly through the crossbar network, which handles any necessary ordering process.
More details about hardware support are explained in Section~\ref{section_5}.

\section{LightNobel Architecture}
\label{section_5}

\begin{figure}[t]
\centering
\includegraphics[width=\colimgwidth]{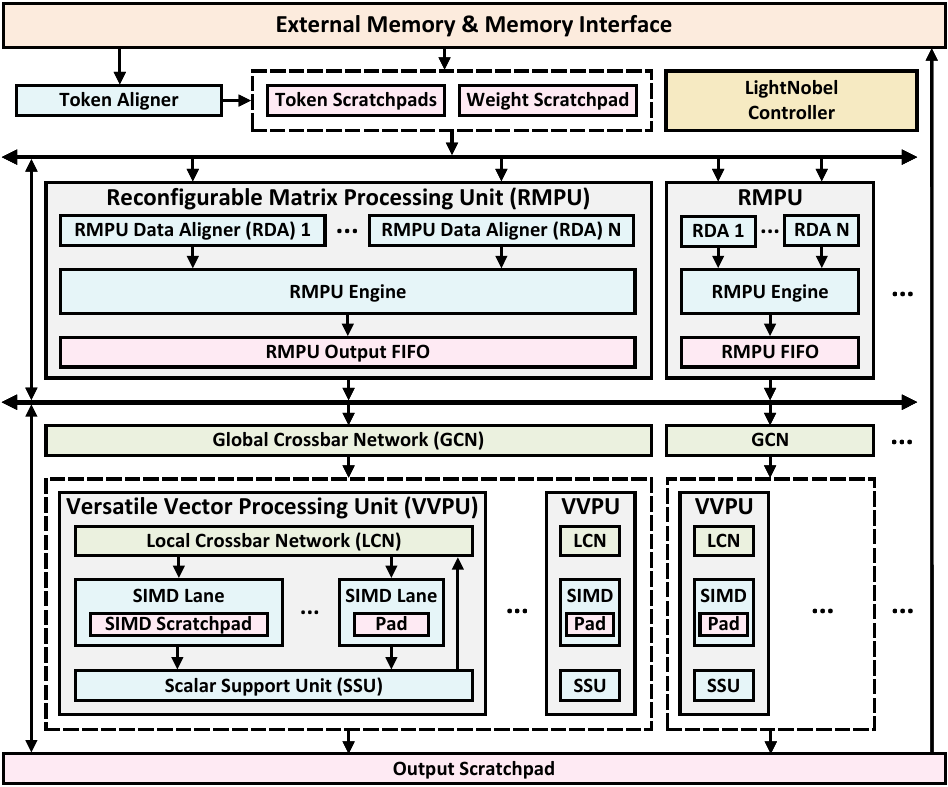}
\caption{Overall block diagram of LightNobel architecture.}
\label{fig_5_1}
\end{figure}

We propose LightNobel, a hardware accelerator designed to efficiently accelerate PPM using the proposed Adaptive Activation Quantization (AAQ). Figure~\ref{fig_5_1} shows the overall architecture of LightNobel. LightNobel is designed to maximize token-level par- allelism by leveraging the benefits of token-wise quantization. To achieve this, it includes a Token Aligner, Reconfigurable Matrix Processing Units (RMPUs), Versatile Vector Processing Units (VVPUs), crossbar data networks, scratchpads, and a controller. A swizzle switch~\cite{swizzle_switch} is employed as a crossbar network to enhance area and power efficiency.

At the beginning of execution, the Token Aligner reads and stores the token block to the Token Scratchpad, which operates in a double-buffering manner to hide memory latency. Simultaneously, weights are preloaded into the Weight Scratchpad for the weight-stationary dataflow that maximizes reuse. The RMPU or VVPU then fetches tokens and weights from these scratchpads.
The RMPU dynamically processes a configurable number of tokens in parallel, adapting to the quantization scheme, while the VVPU executes iterative computations such as LayerNorm. These two units operate in a pipelined manner, where the RMPU passes intermediate results to the VVPU, improving overall throughput.

PPM has a small hidden dimension (e.g., 128) compared to typical attention-based models (e.g., 4,096 in LLaMA 7B~\cite{LLaMA}), leading to small token sizes.
However, the number of tokens is extremely large, easily reaching multi-millions, since it increases quadratically with respect to the protein sequence length. This results in a significant volume of vector operations in the later processing pipelines.
To efficiently manage these workloads, the RMPU is designed for high-throughput token-level operations, while the VVPU specializes in iterative vector computations. 
LightNobel ensures a balanced execution by pairing each RMPU with a fixed number of VVPUs. A Global Crossbar Network (GCN) interconnects all modules and scratchpads, enabling dynamic scheduling between the RMPU and the allocated VVPUs.
For large-scale vector operations such as Softmax or Sequence Representation dataflow, multiple VVPUs can work together as a single large processing unit via the GCN, enhancing their computational capacity.
Finally, computation results are written back to the external main memory via the Output Scratchpad.

\subsection{Token Aligner}

As explained in Section~\ref{section_4_3}, multiple tokens are grouped to enhance memory bandwidth utilization. To allow the RMPU and VVPU to access these blocks efficiently from the scratchpad, the token blocks must be reorganized. This reorganization ensures that each line in the token scratchpad corresponds to the data of a single token, facilitating more efficient token-level processing.
To achieve this, the Token Aligner decodes and realigns the token blocks into a token-wise format before writing them to the scratchpad.
This alignment process ensures seamless execution of subsequent computations by maintaining compatibility with the processing units, maximizing the efficiency of token-wise operations.

\begin{figure*}[t!]
\centering
\includegraphics[width=\fullimgwidth]{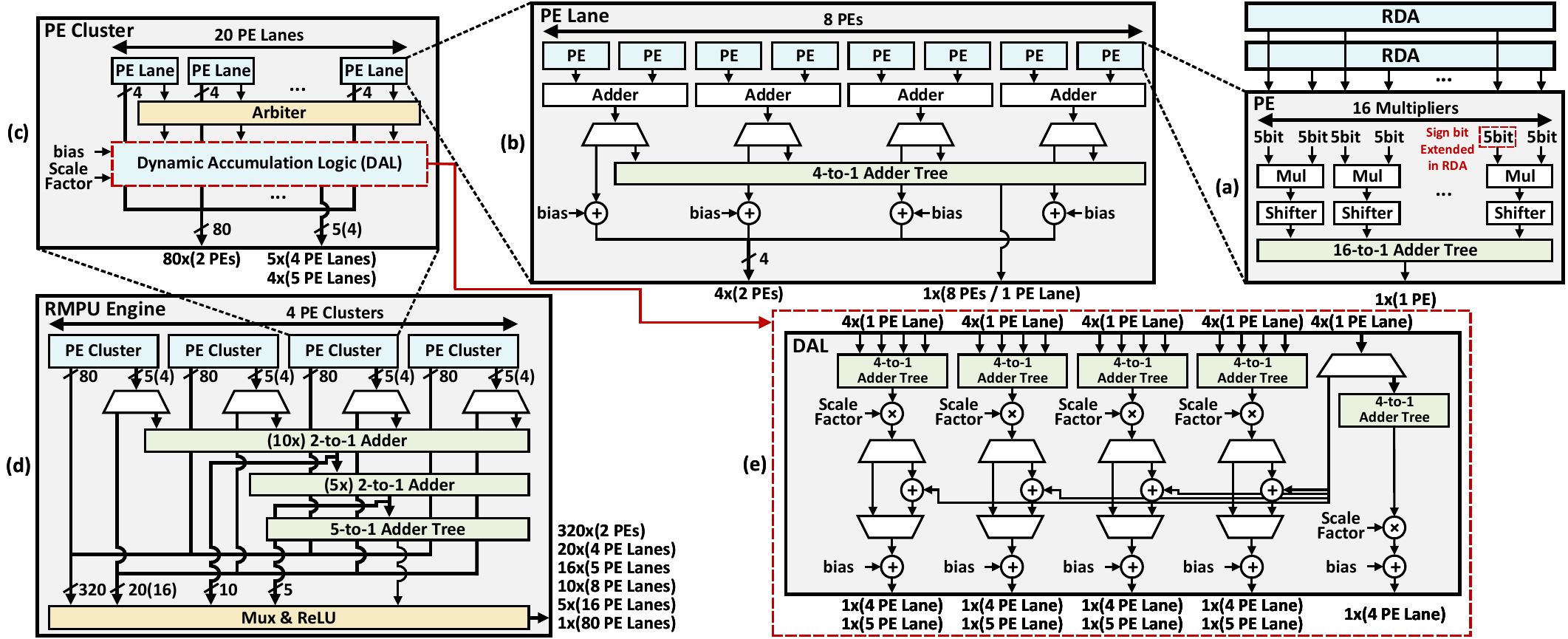}
\caption{Microarchitecture of (a) Processing Element (PE), (b) PE Lane, (c) PE Cluster, (d) RMPU Engine, and (e) Dynamic Accumulation Logic (DAL). For each module, the context provided in parentheses indicates either the source of input or the interpretation of the accumulated output. In the dataflow representation, thin lines denote a single value while thick lines represent multiple values.}
\label{fig_5_3}
\end{figure*}

\subsection{Reconfigurable Matrix Processing Unit}

The Reconfigurable Matrix Processing Unit (RMPU) is designed to efficiently support matrix operations while maximizing computational resource utilization and parallelism. 
A key feature of RMPU is that it can process various data precisions with minimal dequantization. 
Instead of performing dequantization on every piece of data, it minimizes redundant dequantization by prioritizing the execution of operations that can be processed in advance.
However, supporting multi-precision operations introduces another challenge, as the requirement for diverse computational units often leads to severe underutilization. To address this, RMPU employs a bit-level reconfiguration strategy, where data is divided into minimum-unit bit chunks, and computational units are structured to execute numerous small-scale operations. By dynamically allocating computational elements and shifters based on data precision, RMPU ensures efficient hardware utilization~\cite{Bitfusion}.

Despite these advancements, achieving high utilization across multiple data precisions within a single workload still remains a challenge. Without adjusting computational resources at runtime based on operation requirements, underutilization is inevitable. In AAQ, the number of computational units required varies according to the ratio of outliers and the precision level of inliers, even among tokens with the same shape. To address this issue, RMPU introduces a dynamically reconfigurable computation module and a data aligner that optimizes resource allocation based on workload characteristics.

\textbf{RMPU Architecture.} RMPU consists of Reconfigurable Data Aligners (RDAs), RMPU Engine, and RMPU Output FIFOs. The RDA prepares and supplies data according to bit-width requirements, while the RMPU Engine performs computations using a dynamically reconfigurable adder tree structure. The RMPU Output FIFOs efficiently queue and transfer computation results to the GCN. This architecture enables the dynamic allocation of computational resources, maximizing hardware utilization while maintaining high levels of parallelism. Additionally, RMPU leverages token-wise quantization, which eliminates redundant dequantization steps, further improving efficiency.

\textbf{Reconfigurable Data Aligner (RDA).} To ensure efficient computation across varying data precisions, the RMPU Data Aligner (RDA) partitions each token’s data into 4-bit chunks. Here, the 4-bit precision is chosen as it is the lowest that AAQ can optimize in our system, as detailed in Section~\ref{section_4_2} and Section~\ref{section_7_1}. The RDA first splits all input values into 4-bit segments while extracting scaling factors and outlier index information. Subsequently, it sends these segments to the controller for control signal generation.
To maintain consistency with the original data, each chunk undergoes sign extension. Specifically, the chunk containing the most significant bit (MSB) is extended using its MSB value as the sign bit, while all other chunks are extended by appending zeros.
This alignment can fully utilize the RMPU Engine’s computational resources. Although the position of outliers varies across quantized tokens, maintaining a uniform number of 4-bit chunks within tokens that share the same quantization scheme significantly reduces alignment overhead.

\textbf{RMPU Engine.} The RMPU Engine is the core computational unit of RMPU, designed with a dynamic adder tree architecture that allows hierarchical reconfiguration.
Each computational stage produces intermediate results that carry distinct semantic meanings within the processing pipeline. These results are dynamically used across multiple operations, enabling efficient resource allocation and maximizing the utilization of computational units.

At the lowest level, the Processing Engine (PE) serves as the fundamental computation module. As shown in Figure~\ref{fig_5_3}(a), one PE consists of 16 minimal computation units and is capable of performing multiplication between two 16-bit input values.
A PE Lane integrates 8 PEs, as shown in Figure~\ref{fig_5_3}(b), enabling more complex operations such as the inner product. From this stage, the module adopts multiple dataflows to support various operations. Each operation follows a specific dataflow determined by its computational characteristics. The PE Lane supports two dataflows. The first dataflow accumulates the results from two PEs along with a bias, producing 4 output values. It is utilized in the computation process of MHA with a head dimension of 32. The second dataflow employs an adder tree to accumulate the results from all PE modules.

At the next level, the PE Cluster serves as a key computational unit component. As shown in Figure~\ref{fig_5_3}(c), each PE Cluster contains 20 PE Lanes and Dynamic Accumulation Logic (DAL).
A matrix multiplication operation consists of a dot product between two vectors having the same inner dimension. The required number of computational units varies depending on the quantization scheme.
For instance, in a dot product between two 128-dimensional vector tokens, if one token is quantized with 124 inliers at 4-bit precision and four outliers at 16-bit precision while every value in other token is 16-bit precision, the required computational resources are calculated as follows: $4 \times 124\ (\text{inliers}) + 16 \times 4\ (\text{outliers}) = 560$ (4-bit computation units).
Similarly, in PPM operations, most iterations require 4 or 5 PE Lanes for computation. Therefore, the number of PEs is determined to be 20, which is their least common multiple.

However, this introduces two key challenges. First, scaling factors must be considered before accumulation, as inlier multiplication results require scaling before being combined with outlier multiplication results. Second, 4-to-1 and 5-to-1 adder trees are incompatible, necessitating a dynamic approach to accumulation.
To address these challenges, DAL dynamically manages varying numbers of PE Lane outputs while ensuring correct scaling factor application. Figure~\ref{fig_5_3}(e) shows the architecture of DAL. Specifically, for computations requiring 4 PE Lanes, scale factors are applied after accumulation. In contrast, for computations requiring 5 PE Lanes, inlier values are first accumulated and scaled before being combined with the outlier results. To achieve this, one adder tree in DAL is disabled, and the outputs of the remaining four adder trees are accumulated with the result of the fifth PE Lane. The Arbiter is employed at the front of DAL to rearrange PE Lane outputs before entering DAL, ensuring efficient computation.

At the highest level, as shown in Figure~\ref{fig_5_3}(d), the RMPU Engine comprises four PE Clusters.
To accommodate various workloads, the RMPU Engine produces multiple types of output results.
The sum of the 2 PE results corresponds to the output required for MHA with a head dimension of 32, as described earlier.
The sums of 4 and 5 PE Lane results serve as computational outputs for quantized tokens processed within the DAL, and the sums of 8 or 16 PE Lane results serve as computational outputs for non-quantized tokens.
Additionally, the engine generates the sum of 80 PE results, enabling scalability for larger computations that require multiple RMPUs.
As the last layer, it also performs ReLU operations.
Ultimately, a single RMPU Engine supports up to 20 tokens simultaneously, achieving a high level of token parallelism.

\subsection{Versatile Vector Processing Unit}
The Versatile Vector Processing Unit (VVPU) is designed to support all vector operations required for PPM, including LayerNorm, Softmax, residual connections, and the new operations required for AAQ.
By having a unified structure, the VVPU eliminates the need for separate dedicated components for each operation, achieving both high resource utilization and operational flexibility.

\begin{figure}[t]
\centering
\includegraphics[width=\colimgwidth]{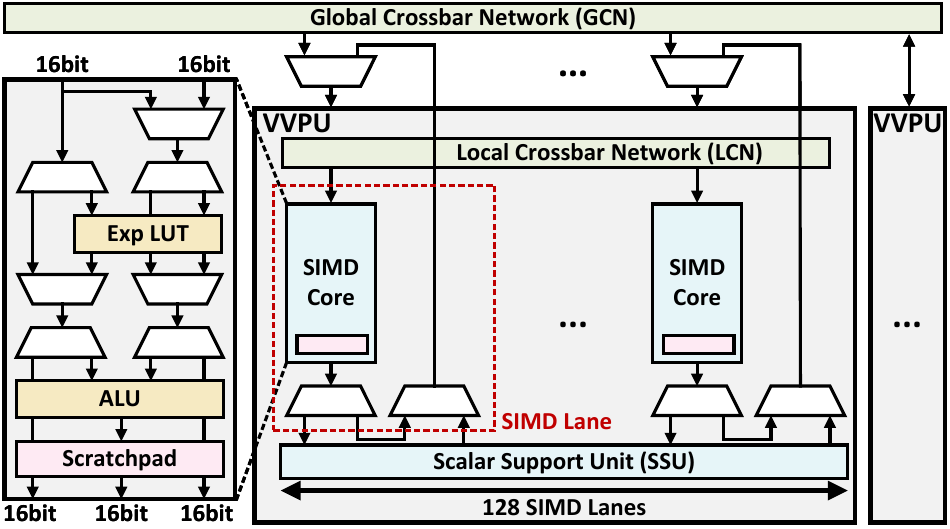}
\caption{Microarchitecture of VVPU.}
\label{fig_5_4}
\end{figure}

Figure~\ref{fig_5_4} shows the microarchitecture of VVPU.
The VVPU comprises multiple SIMD Lanes, a Scalar Support Unit (SSU), and a Local Crossbar Network (LCN). Each SIMD Lane includes a SIMD Core with an ALU that can process operations between two 16-bit operands for weights, scratchpad memory, and a two-level exponent lookup table~\cite{A3}. 
These SIMD Lanes are interconnected via the LCN, which allows it to handle data alignment problems dynamically during runtime.
SSU enhances the overall efficiency of the VVPU by handling scalar operations such as averaging and data formatting for quantization tokens. This offloading mechanism ensures that the SIMD Lanes remain dedicated to higher-complexity computations, optimizing the utilization of hardware resources.

\textbf{Dynamic Top-k Selection.} To determine varying numbers of outliers in the quantization scheme, identifying the top-k values at runtime, the k largest elements among all token values, is essential. 
The VVPU addresses this requirement by leveraging hardware parallelism on bitonic top-k sorting~\cite{bitonic_topk}.
During the sorting process, the indices of the values are continuously tracked, enabling the controller to identify the locations of the top-k values in the final stage. 
This approach eliminates the need for additional sorting modules, making the VVPU highly efficient for top-k operations required in quantization. 
Moreover, when configured with k=1, the VVPU can search the maximum in operations such as Softmax.

\textbf{Runtime Quantization.} One of the key functionalities of the VVPU is the runtime quantization. 
Unlike weight, activations are dynamically generated during runtime. 
Therefore, runtime quantization is critical for reducing memory footprint and computational requirements while enabling the continuation of operations.
The quantization process begins with the top-k operation, where the VVPU identifies outliers and scaling factors using SIMD lanes. 
Each value is then scaled by the identified scaling factor. Then, LCN reorders the quantized values according to the memory layout for quantized data.
Finally, the SSU aligns the necessary values to conform to the memory layout. 

\begin{figure*}[t]
\centering
\includegraphics[width=\fullimgwidth]{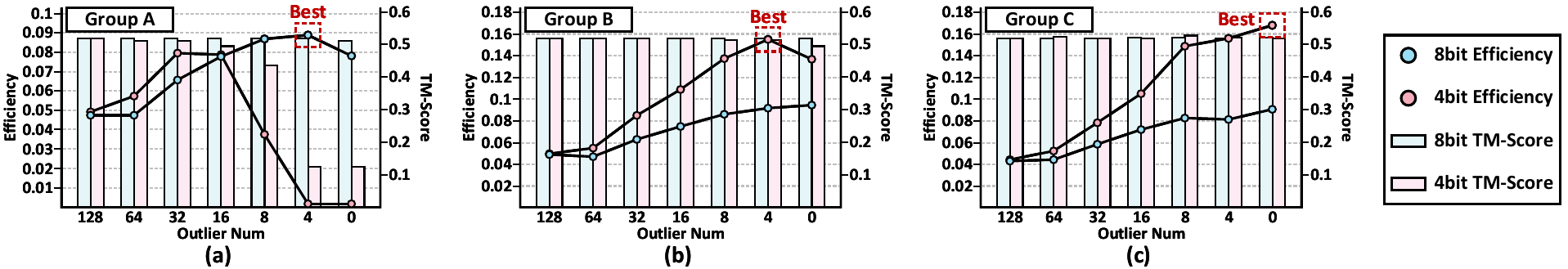}
\caption{Design space exploration on quantization scheme for (a) Group A, (b) Group B, and (c) Group C.}
\label{fig_7_1}
\end{figure*}

\subsection{Token-wise Multi-head Attention} 

LightNobel is designed to maximize the efficiency of PPM by leveraging token-wise dataflow. To align with this design principle, the MHA mechanism is also implemented using token-wise dataflow. During the MHA computation, LightNobel eliminates the need for writeback and read of intermediate activations, such as the attention score matrix. This approach is similar to FlashAttention~\cite{FlashAttention} but with optimizations tailored for token-wise operation.

The RMPU first performs a multiplication operation between Q and K in parallel for each head, while the VVPU handles dequantization and accumulation of intermediate results simultaneously. 
This computation is repeated iteratively, and the VVPU applies the softmax operation to the outputs, pipelined with the Q and K multiplication steps, minimizing latency in finding the maximum value. 
Finally, the V tokens are loaded, multiplied by the softmax results, and written to memory, completing the MHA computation. 
This process significantly alleviates the high peak memory requirement that arises when storing the entire score matrix.

In conventional attention-based models, token-wise approaches are challenging due to the limited opportunities for memory reuse due to large hidden dimensions.
However, our system benefits from PPM having a very small hidden dimension. Since PPM has a significantly large total score matrix size due to the activation dimension, not offloading intermediate values achieves considerable gains not only in memory space requirements but also in terms of memory footprint.

\section{Methodology}
\label{section_6}

\textbf{System Evaluation Method.} To measure the performance of the designed system, we implement a cycle-accurate simulator in Python~\cite{Python}.
Also, we implement all the logic modules of the system in RTL using SystemVerilog~\cite{SystemVerilog} and synthesize using Synopsys Design Compiler~\cite{Synopsys_Design_Compiler} 2019.3 at 28nm technology. The synthesis is performed targeting a 1 GHz operating frequency. For on-chip scratchpads, we estimate power and area using a 28nm memory compiler and Cacti 7.0~\cite{Cacti}. Since Cacti 7.0 only supports up to 32nm technology, we carefully downscale it to 28nm technology using scaling factors as done in existing papers~\cite{cacti_down_1, cacti_down_2, cacti_down_3, BLESS}.

The overall latency of LightNobel is determined by the summation of the longest delay of each pipelining stage.
LightNobel's key stages include RMPU operations (e.g., Linear), VVPU operations (e.g., quantization), and memory operations (e.g., data read/write).
For cycle-accurate simulation, we first ensure that the latency of key modules is accurately modeled at every stage. Since the RMPU is based on a MAC tree architecture, we evaluate its throughput. Meanwhile, as the VVPU executes iterative operations, we use additional C-based~\cite{C} control signal simulation to measure its latency. Also, we use Ramulator~\cite{Ramulator} to accurately simulate memory operations, considering data bus width and burst length alignment. Here, we use 80 GB of 5 HBM2E memory stacks~\cite{HBM2E} for fair comparisons with the baseline GPUs~\cite{A100, H100}. 
To ensure the reliability of the Python-based simulator, we cross-validate the simulation results of modules against the RTL-based simulation results.
The cross-validation on CAMEO~\cite{CAMEO}, CASP14~\cite{CASP14}, CASP15~\cite{CASP15}, and CASP16~\cite{CASP16} datasets shows the discrepancies of 4.63\%, 3.62\%, 3.14\%, and 1.81\%, respectively, averaging 3.30\%. These differences mainly arise from the tail latency of each stage, which decreases as the sequence length increases. Consequently, the overall discrepancy remains within 5\% for any cases, demonstrating that the Python-based simulator is reliable with a permissible error rate.

\textbf{Datasets.} For performance evaluation, we use CAMEO~\cite{CAMEO}, CA- SP14~\cite{CASP14}, CASP15~\cite{CASP15}, and CASP16~\cite{CASP16} datasets.
These datasets evaluate the predicted structures of proteins against experimentally determined results provided by the PDB~\cite{PDB} and are widely recognized as the standard benchmarks in the field of protein structure prediction.
For CASP16, since the competition is still ongoing, the ground truth data has not yet been released. Hence, accuracy evaluation is conducted on datasets excluding CASP16, while other performance metrics are evaluated across all datasets.

\textbf{Baseline Comparison Method.} As the baseline PPM for evaluation, we use ESMFold (Commit 2b36991)~\cite{ESM_GIT}. Although AlphaFold2~\cite{AlphaFold2} and ESMFold~\cite{ESMFold} differ slightly in Sequence Representation dataflow, they entirely share the same Pair Representation dataflow, which is the main focus of our paper.
Thus, we select ESMFold as the baseline PPM due to its faster speed and simpler structure.
As an Input Embedding model, we use ESM-2 model with 3 billion parameters (esm2\_t36\_3B\_UR50D) as the protein language model.
For the chunk option, we employ the Chunk4 option, consistent with the configuration used in the AlphaFold2~\cite{AlphaFold2}.
Since there is no existing hardware accelerator work targeting PPM, we compare our system against the latest GPUs as a hardware baseline.
To evaluate PPMs, we utilize a Linux server environment with two Intel(R) Xeon(R) Platinum 8452Y CPU (36 core/72 thread) operating at 2.90GHz~\cite{CPU}, 1TB DDR5 memory, and GPUs including NVIDIA H100 80GB PCIe (in short, H100)~\cite{H100} and NVIDIA A100 80GB PCIe (in short, A100)~\cite{A100}. To analyze the GPU execution, we use the NVIDIA Nsight Systems~\cite{Nsight_sys}.

\section{Design Space Exploration}
\label{section_7}

\subsection{AAQ Quantization Scheme}
\label{section_7_1}

\begin{figure}[t]
\centering
\includegraphics[width=\colimgwidth]{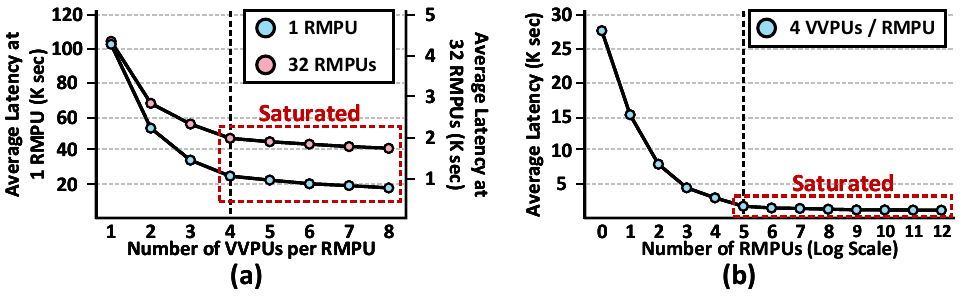}
\caption{Design space exploration on hardware configuration with respect to (a) the number of VVPUs per RMPU and (b) the total number of RMPUs.}
\label{fig_7_2}
\vspace{-0.10in}
\end{figure}

We conduct design space exploration on the efficiency and TM-Score changes across various quantization schemes to identify the optimal point in the AAQ algorithm.
Figure~\ref{fig_7_1} shows the efficiency and TM-Score variations across Groups A, B, and C, as described in Section~\ref{section_4_2}.
Since accuracy evaluation requires ground truth data, we use CAMEO~\cite{CAMEO}, CASP14~\cite{CASP14}, and CASP15~\cite{CASP15} datasets for the experiment.
Efficiency is calculated by considering the memory size of the quantized tokens and the resulting TM-Score for each configuration, while it decreases significantly as TM-Score drops, targeting to minimize accuracy degradation.

In Group A, Figure~\ref{fig_7_1}(a) shows that the quantization using 8-bit precision for inliers with 4 outliers handling achieves the best efficiency. 
When using 4-bit precision for inliers, handling fewer than 32 outliers reduces TM-Score, while handling more increases quantized token size, lowering efficiency. 
When using 8-bit precision for inliers, handling at least 4 outliers prevents TM-Score drops.
In Group B, Figure~\ref{fig_7_1}(b) shows that the quantization using 4-bit precision for inliers with 4 outliers handling achieves the best efficiency.
With 4-bit precision for inliers, handling fewer than 4 outliers degrades the TM-Score, but handling 4 or more prevents TM-Score drops. With 8-bit precision for inliers, the TM-Score remains stable, but the quantized token size increases.
In Group C, Figure~\ref{fig_7_1}(c) shows that the quantization using 4-bit precision for inliers without outlier handling achieves the best efficiency. 
TM-Score remains stable across all configurations, regardless of inlier precision or even without outlier handling. Thus, the smallest is the most efficient.

\subsection{Hardware Configuration}

We conduct design space exploration on the performance variations on LightNobel hardware configuration to identify the optimal point of hardware design.
Figure~\ref{fig_7_2}(a) shows the average latency of PPM as the number of VVPUs per RMPU varies, conducted with a single RMPU and 32 RMPUs to assess the contribution of VVPUs to the overall performance. 
In both cases, the latency saturates at 4 VVPUs per RMPU. 
This saturation is attributed to the high token parallelism of RMPUs and the small hidden dimensions, which limit the operations executed by VVPUs to being hidden when the number of VVPUs is small.
With a single RMPU, latency gradually decreases without saturation as the number of RMPUs increases. This degradation is due to the overall latency being dominated by the VVPU operation time, as there are only a small number of VVPUs in the system.

Figure~\ref{fig_7_2}(b) shows the average latency of PPM as the number of RMPUs varies when the number of VVPUs per RMPU is fixed to 4. The result shows that the performance saturates at 32 RMPUs. This saturation is because having 32 RMPUs provides sufficient computational resources to process data fetched from memory. Adding more than 32 RMPUs slightly improves performance by increasing the number of VVPUs. However, the performance gains are minimal compared to the increase in RMPU numbers, which would not justify the additional area and power overhead.

\section{Experimental Results}
\label {section_8}

\subsection{Accuracy Evaluation}
\label {section_8_1}

For algorithmic validation, we evaluate the accuracy of AAQ with various recent quantization schemes.
Table~\ref{table_8_1_1} summarizes the properties of targeting quantization schemes, including AAQ, providing details on the activation memory footprint, weight memory size, and total memory footprint when they are applied to PPM with the longest protein in CASP15 dataset which has sequence length (amino acids) of 3,364 (T1169).
Although LightNobel further reduces the activation memory footprint via token-wise MHA, we exclude this hardware-driven advantage for a fair comparison. We also conduct evaluation solely on parts that share the same dataflow among quantization schemes.

\begin{table}[t]
\caption{Description of various quantization schemes.}
\centering
\includegraphics[width=\colimgwidth]{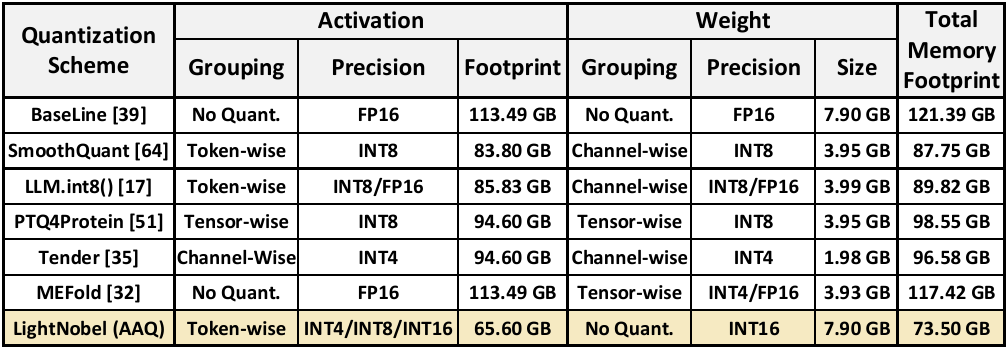}
\label{table_8_1_1}
\vspace{-0.061in}
\end{table}

\begin{figure}[t]
\centering
\includegraphics[width=\colimgwidth]{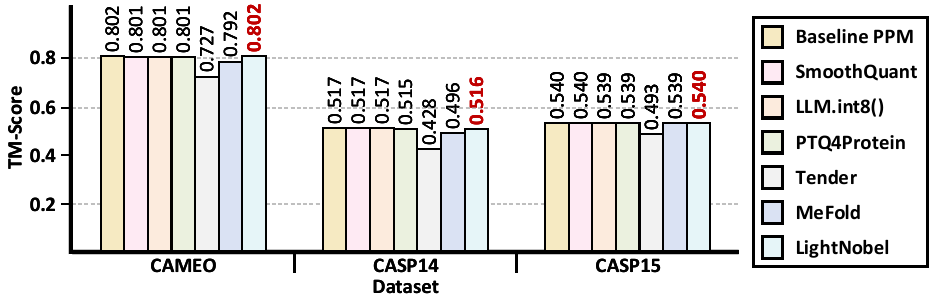}
\caption{Accuracy evaluation result across datasets.}
\label{fig_8_1_2} 
\end{figure}

Figure~\ref{fig_8_1_2} shows the average TM-Scores across different datasets when the quantization schemes are applied to PPM.
Tender~\cite{Tender} and MeFold~\cite{MeFold} significantly degraded the TM-Score. This drop indicates that additional solutions are necessary to achieve quantization below INT8 precision in PPM without compromising accuracy.
Other quantization schemes exhibited acceptable TM-Score variations, keeping a loss below 0.002. However, due to their use of high-precision quantization, schemes other than LightNobel incurred a relatively higher total memory footprint.
AAQ in LightNobel achieved a negligible TM-Score change of less than 0.001 while maintaining a minimum total memory footprint.
A TM-Score above 0.5 signifies meaningful prediction results, which confirms that our approach achieves significant protein modeling outcomes. This advantage is attributed to AAQ’s capability to adapt the quantization scheme to the unique characteristics of each activation.

\subsection{Performance Evaluation}
\label {section_8_2}

\textbf{End-to-end PPM Model Performance.} We evaluate the end-to-end performance of various recent PPMs, including LightNobel.
For our experiments, we use proteins with sequence lengths of less than 1,410 that fit within an 80 GB memory constraint from the CASP16 dataset.
All models except LightNobel are evaluated on H100 using the vanilla option.
Since LightNobel accelerates the Protein Folding Block and is assumed to operate with a CPU, we equalize its data transfer latency with the baseline for a fair comparison.

\begin{figure*}[t]
\centering
\footnotesize
\includegraphics[width=\fullimgwidth]{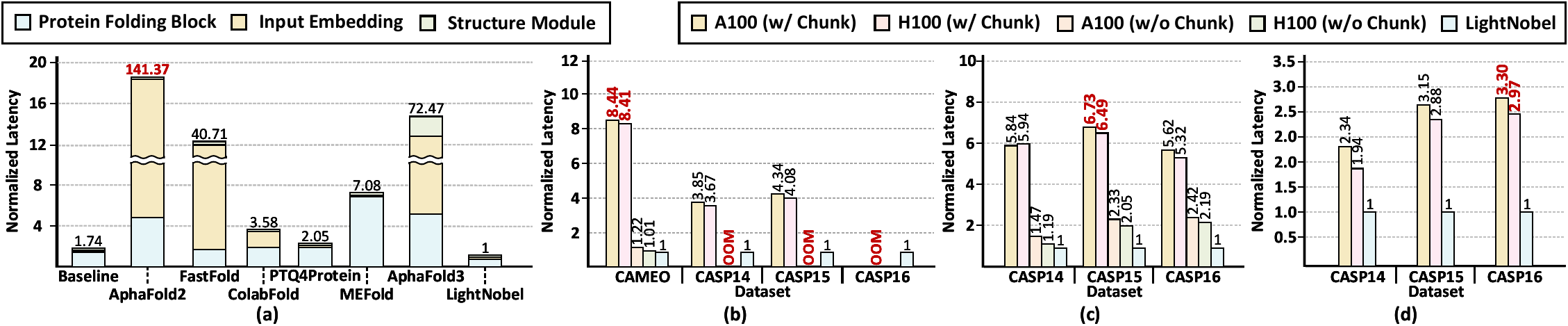}
\caption{(a) End-to-end performance evaluation result across various recent PPMs, and hardware performance evaluation result across datasets with (b) all proteins, (c) proteins excluding those that incur OOMs, and (d) proteins that can only be executed with the chunk option.}
\label{fig_8_2} 
\vspace{-0.01in}
\end{figure*}

Figure~\ref{fig_8_2}(a) shows the normalized end-to-end performance across various PPMs.
LightNobel outperforms the least-performing model, MeFold~\cite{MeFold}, by 8.22$\times$ and even outperforms the best-performing model, ESMFold~\cite{ESMFold}, by 1.11$\times$ in Protein Folding Block performance.
This result demonstrates that LightNobel effectively addresses memory overhead issues with minimum computational resources.
For end-to-end performance, LightNobel also outperforms the least-performing model, AlphaFold2~\cite{AlphaFold2}, by 141.37$\times$ and the best-performing model, ESMFold~\cite{ESMFold}, by 1.74$\times$.
AlphaFold2~\cite{AlphaFold2}, FastFold~\cite{FastFold}, ColabFold~\cite{ColabFold}, and AlphaFold3~\cite{AlphaFold3} suffer from long Input Embedding times due to database search, especially for long sequences.
While FastFold~\cite{FastFold} and ColabFold~\cite{ColabFold} attempt to address this issue, it still remains a bottleneck, highlighting ESMFold~\cite{ESMFold} as a strong baseline.
Among the models that use protein language models for Input Embedding, including ESMFold~\cite{ESMFold}, ColabFold~\cite{ColabFold}, PTQ4Protein~\cite{PTQ4Protein}, and MEFold~\cite{MeFold}, LightNobel archives the best performance.
This result is attributed to LightNobel's superior acceleration of the Protein Folding Block, which is a major bottleneck of the overall latency.

\textbf{Hardware Performance.} We evaluate the performance of LightNobel hardware and NVIDIA A100 and H100 GPUs, focusing on the Protein Folding Block.
We use CAMEO, CASP14, CASP15, and CASP16 datasets for the experiment.
Figure~\ref{fig_8_2}(b) shows the normalized latency across datasets.
LightNobel achieves 3.85-8.44$\times$, 3.67-8.41$\times$ lower latency with the chunk option and 1.22$\times$, 1.01$\times$ lower latency without the chunk option compared to A100 and H100.
The chunk option significantly increases GPU latency due to kernel overhead from frequent kernel calls and returns, highlighting LightNobel’s advantage in handling long sequence lengths.
Moreover, despite H100’s 5$\times$ higher INT8 computing resources compared to A100 (e.g., 3,026 TOPS vs. 624 TOPS), performance gains remain minimal due to the large portion of the PPM workload being memory-bounded, leading to low utilization of compute resources~\cite{Scalefold}.
Despite LightNobel having only 537 TOPS of computational resources, it demonstrates significantly better performance compared to A100 and H100 under the same 2TB/s bandwidth.
These results demonstrate the superior performance efficiency of LightNobel and suggest that similar trends will be observed with the NVIDIA H200, the state-of-the-art GPU~\cite{H200}.

In experiments across the entire dataset, GPUs face out-of-mem- ory (OOM) issues.
Therefore, for a fair comparison, we exclude the proteins that cannot be processed on GPUs without the chunk option and conduct experiments on the remaining datasets. The CAMEO dataset is excluded because it can already be fully processed without the chunk option.
Figure~\ref{fig_8_2}(c) shows LightNobel achieved 5.62-6.73$\times$, 5.32-6.49$\times$ lower latency with the chunk option and 1.47-2.42$\times$, 1.19-2.19$\times$ lower latency without the chunk option compared to A100 and H100 in this experiment.
We also conduct experiments on proteins that GPUs cannot process without the chunk option to evaluate the performance of LightNobel on proteins with long sequence lengths.
Figure~\ref{fig_8_2}(d) shows LightNobel achieves 2.34-3.30$\times$, 1.94-2.97$\times$ lower latency with the chunk option compared to A100 and H100 in this experiment.
For short proteins, the kernel overhead constitutes a significant portion of the overall latency, leading to relatively large speedup gains. However, as the sequence length increases, this overhead becomes less dominant. Although the absolute speedup is relatively modest, LightNobel’s speedup becomes more stable and consistent, demonstrating a high degree of scalability with respect to sequence length.

\begin{figure}[t]
\centering
\includegraphics[width=\colimgwidth]{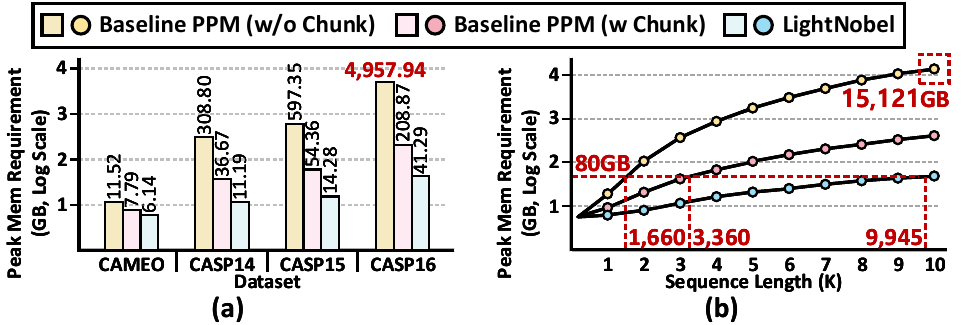}
\caption{Peak memory requirement of PPM across (a) datasets and (b) various sequence lengths.}
\label{fig_8_3} 
\vspace{-0.10in}
\end{figure}

\begin{figure}[t]
\centering
\includegraphics[width=\colimgwidth]{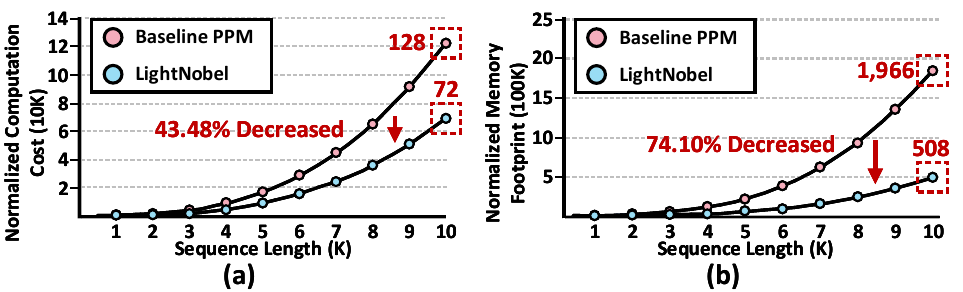}
\caption{(a) Computational cost of PPM and (b) memory footprint of PPM across various sequence lengths.}
\label{fig_8_4} 
\vspace{-0.10in}
\end{figure}

\subsection{In-Depth Analysis}

\textbf{Peak Memory Requirement.} To evaluate LightNobel's benefit on peak memory requirement, we measure the peak memory requirements across various datasets. 
Figure~\ref{fig_8_3}(a) shows the peak memory requirement of baseline PPM and LightNobel. 
LightNobel achieves 1.87-120.05$\times$ lower peak memory requirement without the chunk options and 1.26-5.05$\times$ lower requirements with the chunk option compared to the baseline PPM.
For more detailed analysis, we also measure peak memory requirement across varying sequence lengths. Figure~\ref{fig_8_3}(b) shows the peak memory requirements as the protein's sequence length increases. Due to memory limitations, running every protein on 80 GB of GPU VRAM is infeasible. Therefore, we measure peak memory requirements through actual GPU executions for shorter protein sequences and estimate the requirements by applying equivalent computational processes for longer protein sequences. 
Without the chunk option, GPUs inefficiently store multiple intermediate activations, such as the score matrix, leading to high peak memory requirements. The chunk option mitigates this overhead by processing data in smaller chunks but still retains redundant intermediate activations in memory, resulting in non-negligible memory overhead.
LightNobel significantly reduces peak memory requirements by employing quantization and Token-wise Multi-Head Attention, enabling computation at the token level. AAQ compresses activations, further minimizing memory usage. Unlike chunking, which processes data channel-wise, LightNobel achieves higher efficiency through parallel token-level computation. As a result, LightNobel processes every dataset within 80 GB of memory, supporting sequence lengths of up to 9,945, which is 1.45$\times$ longer than the longest protein in CASP16 dataset, which is 6,879.

\textbf{Computational Benefits.} To evaluate LightNobel's benefit on computational cost, we conduct experiments comparing the computational costs of the baseline PPM and LightNobel. 
We evaluate LightNobel’s computational benefits by comparing its computational cost with the baseline PPM.
Figure~\ref{fig_8_4}(a) shows the computational cost as sequence length increases. To calculate the computational cost, we convert every operation to equivalent INT8 operations and accumulate.
LightNobel reduces the average computational cost by 43.38\% compared to baseline PPM due to two key factors.
First, AAQ lowers the cost of single data computation, particularly for multiplications, which scale quadratically with precision reduction. 
Second, LightNobel eliminates redundant dequantization in matrix multiplications by applying the scale factor only once at the end rather than repeatedly for each value, optimizing the most compute-intensive operation in attention-based models.

\textbf{Memory Footprint Benefits.} To evaluate LightNobel's benefits on memory footprint, we conduct experiments comparing the memory footprints of the baseline PPM and LightNobel. 
Figure~\ref{fig_8_4}(b) shows the memory footprint as the sequence length increases. 
LightNobel achieves 74.10\% lower memory footprint on average. 
This reduction stems from AAQ, which minimizes activation size via quantization during each PPM operation.
Additionally, LightNobel quantizes residual connections between layers, often overlooked in prior studies, further enabling a smaller memory footprint and improving scalability.
Since the number of tokens increases quadratically with sequence length in PPM, token-wise quantization proportionally reduces peak memory usage, enabling efficient processing of longer sequences.
This property ensures that LightNobel can efficiently handle longer sequences.

\begin{table}[t]
\caption{Area and power analysis of LightNobel.}
\centering
\includegraphics[width=\colimgwidth]{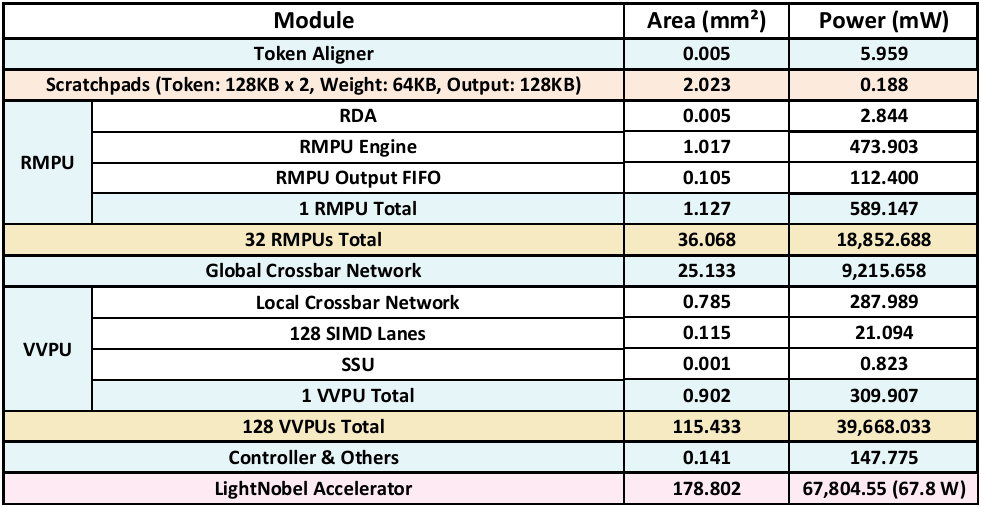}
\label{table_8_5}
\vspace{-0.20in}
\end{table}

\subsection{Area and Power Analysis}

We conduct area and power analysis of the proposed system. Table~\ref{table_8_5} shows the detailed breakdown of the area and power estimations of the LightNobel accelerator. The total area is 178.80 mm\textsuperscript{2}, and the total power consumption is 67.80 W.
The crossbar networks are the most dominant component, accounting for 70.28\% of the total area and 67.95\% of the total power consumption.
The second dominant component is the RMPU Engine, which accounts for 18.20\% of the total area and 22.36\% of the total power consumption.
The crossbar networks play a pivotal role in the system’s dataflow management, including pipelining and ordering for dynamic dataflow, enabling dequantization-free computation for multi-precision values in the RMPU as well as top-k sorting and quantization in the VVPU. Also, the RMPU Engine serves as the core module for computational power. These results justify the observed area and power consumption of LightNobel.

When compared to GPUs, LightNobel requires only 21.94\% of area and 19.37\% of power compared to A100, and 21.63\% of area and 22.60\% of power compared to H100.
It achieves up to 37.29$\times$, 43.35$\times$ higher power efficiency than A100 and H100 with the chunk option and up to 5.39$\times$, 5.21$\times$ without it. 
These results are particularly significant, as the LightNobel accelerator is implemented in a 28nm process, whereas A100 and H100 use more advanced 7nm and 4nm processes, underscoring LightNobel’s superior area and power efficiency.
Moreover, since LightNobel supports significantly longer sequence lengths compared to GPUs, it is expected to have even better efficiency for longer sequence lengths.

\section{Related Works and Discussion}
\label{section_9}

\subsection{Previous Works on PPM}
\label {section_9_1}

There have been efforts to optimize PPM, but they have failed to address critical memory-related challenges.
Fastfold~\cite{FastFold} and Scalefold~\cite{Scalefold} tackle communication overhead issues between multiple GPUs during PPM training by employing scheduling and parallelism techniques. While these methods improve training scalability, the benefits are limited in the inference phase. 
MEFold~\cite{MeFold} and PTQ4Protein~\cite{PTQ4Protein} introduce quantization to PPM.
MEFold applies weight-only quantization, leaving memory-related challenges arising from activation unresolved. It supports a maximum sequence length of 2,828 with a peak memory requirement of 78.7 GB in an 80 GB memory environment. LightNobel achieves the same with just 12.1 GB of memory, improving scalability by 6.05$\times$.
PTQ4Protein quantizes both weights and activations but conducts experiments only on proteins with a maximum sequence length of 700, reducing peak memory to 11.6 GB. For the same sequence length, LightNobel achieves a peak memory usage of 7.1 GB, indicating 1.63$\times$ better scalability. These gaps widen with longer sequences.
Moreover, both Mefold and PTQ4Protein suffer from significant accuracy degradation as their quantization precision decreases, which can pose critical issues in biological modeling. LightNobel mitigates these challenges by applying AAQ.

\subsection{Quantization for Attention-based Models}
\label {section_9_2}

Prior works, such as SmoothQuant~\cite{SmoothQuant}, Qserve~\cite{QServe}, and AWQ~\cite{AWQ}, suggest efficient GPU-based quantization algorithms.
These methods handle outliers by leveraging their characteristics, separating them to enable low-precision computations with minimal loss of accuracy.
However, they apply the same precision to all values except outliers due to limitations of conventional hardware, such as GPUs, which limit the quantization performance. 
Mokey~\cite{Mokey}, Olive~\cite{Olive}, and Tender~\cite{Tender} adopt additional hardware for more aggressive quantization. They propose accelerators alongside quantization schemes to achieve high accuracy. 
However, these methods are not suitable for PPM since they cannot exploit the distinct characteristics of data values, which can lead to accuracy degradation.

\subsection{AAQ Challenges on Existing Hardware}
\label {section_9_3}

\textbf{Challenges of AAQ on GPU.} The performance of GPU kernels depends on how efficiently the MMA performance of Tensor cores is used~\cite{QServe}. However, activation quantization is inefficient on GPUs due to their heavy reliance on CUDA Cores rather than Tensor Cores because activation quantization requires runtime dequantization and quantization, unlike weight quantization. Actually, W4A4 quantization (e.g., QuaRot~\cite{QuaRot}) is slower than FP16 execution in TensorRT-LLM~\cite{QServe}. In AAQ, dynamic precision and outlier handling further increase reliance on CUDA Cores.
Additionally, multi-precision execution suffers from warp divergence in SIMD and SIMT architectures, reducing utilization.

\textbf{Challenges of AAQ on Existing Accelerator.} Existing accelerators for attention-based models, such as Mokey~\cite{Mokey} and Olive~\cite{Olive}, are optimized for tensor-wise quantization with efficient memory layouts, while Tender~\cite{Tender} is designed for channel-wise quantization, leveraging shifters for efficient dequantization and runtime quantization.
However, AAQ requires token-wise quantization, where each token has a distinct scaling factor, and outliers are handled dynamically at runtime. This requirement increases memory overhead and leads to redundant dequantization on existing hardware, ultimately eliminating the advantages of memory layout and architecture in existing accelerators.
Also, Olive~\cite{Olive} does not have a hardware encoder for runtime quantization. 
Moreover, existing accelerators do not support multi-precision operations or dynamic dataflows, making it challenging to maintain high utilization.
Additionally, top-k operations required by AAQ are not natively supported. 
Consequently, dedicated hardware is necessary for the efficient execution of AAQ.

\section{Conclusion}

This paper presents LightNobel, a hardware-software co-designed solution that addresses the scalability limitations in sequence length for the Protein Structure Prediction Model (PPM) caused by excessive activation size.
We propose Token-wise Adaptive Activation Quantization (AAQ), a quantization method that significantly reduces activation size without compromising prediction accuracy, leading to substantial reductions in memory requirements and computational costs.
Our hardware innovations, including the Reconfigurable Matrix Processing Unit (RMPU) and the Versatile Vector Processing Unit (VVPU), enable efficient handling of dynamically quantized multi-precision data in token-wise dataflow, pushing the boundaries of hardware utilization and computational efficiency for AAQ support.
LightNobel achieves up to 8.44$\times$, 8.41$\times$ speedup and 37.29$\times$, 43.35$\times$ higher power efficiency over the latest NVIDIA A100 and H100 GPUs, respectively, while maintaining negligible accuracy loss. It also reduces the peak memory requirement up to 120.05$\times$, enabling scalable processing for proteins with long sequences.
These results demonstrate that LightNobel offers a highly scalable and high-performance solution for PPM, laying the groundwork for next-generation protein structure prediction accelerators.

\begin{acks}
This work was partly supported by Institute of Information \& communications Technology Planning \& Evaluation (IITP) grant funded by the Korea government (MSIT) (No. 2022-0-01036, Development of Ultra-Performance PIM Processor Soc with PFLOPS-Performance and GByte-Memory) and (No.2022-0-01037, Development of High Performance Processing-In-Memory Technology based on DRAM), and Samsung Electronics Co., Ltd.
\end{acks}

\bibliographystyle{ACM-Reference-Format}
\bibliography{refs}

\end{document}